\documentclass[%
 reprint,
 amsmath,amssymb,
 aps,
]{revtex4-1}

\usepackage{graphicx}
\usepackage{dcolumn}
\usepackage{bm}

\begin{document}

\preprint{APS/123-QED}

\title{First measurement of several $\beta$-delayed neutron emitting isotopes beyond N=126}

\author{R.~Caballero-Folch$^{1}$, C.~Domingo-Pardo$^{2,*}$, J.~Agramunt$^{2}$,
  A.~Algora$^{2,3}$, F.~Ameil$^{4}$, A.~Arcones$^{4}$, Y.~Ayyad$^{5}$, J.~Benlliure$^{5}$, I.N.~Borzov$^{6,7}$,
  M.~Bowry$^{8}$, F.~Calvi\~no$^{1}$, D.~Cano-Ott$^{9}$, G.~Cort\'es$^{1}$, T.~Davinson$^{10}$,
  I.~Dillmann$^{4,11}$, A.~Estrade$^{4,12}$, A.~Evdokimov$^{4,11}$,
  T.~Faestermann$^{13}$, F.~Farinon$^{4}$, D.~Galaviz$^{14}$,
  A.R.~Garc\'ia$^{9}$, H.~Geissel$^{4,11}$, W.~Gelletly$^{8}$,
  R.~Gernh\"auser$^{13}$, M.B.~G\'omez-Hornillos$^{1}$, C.~Guerrero$^{15}$,
  M.~Heil$^{4}$, C.~Hinke$^{13}$, R.~Kn\"obel$^{4}$, I.~Kojouharov$^{4}$,
  J.~Kurcewicz$^{4}$, N.~Kurz$^{4}$, Y.~Litvinov$^{4}$, L.~Maier$^{13}$,
  J.~Marganiec$^{16}$, T.~Marketin$^{17}$, M.~Marta$^{4}$, T.~Mart\'inez$^{9}$, G.~Mart\'inez-Pinedo$^{4}$, F.~Montes$^{18,19}$, I.~Mukha$^{4}$, D.R.~Napoli$^{20}$,
  C.~Nociforo$^{4}$, C.~Paradela$^{5}$, S.~Pietri$^{4}$,
  Zs.~Podoly\'ak$^{8}$, A.~Prochazka$^{4}$, S.~Rice$^{8}$, A.~Riego$^{1}$,
  B.~Rubio$^{2}$, H.~Schaffner$^{4}$, Ch.~Scheidenberger$^{4,11}$,
  K.~Smith$^{4,17,19,21,22}$, E.~Sokol$^{23}$, K.~Steiger$^{13}$, B.~Sun$^{4}$,
  J.L.~Ta\'in$^{2}$, M.~Takechi$^{4}$, D.~Testov$^{23,24}$, H.~Weick$^{4}$,
  E.~Wilson$^{8}$, J.S.~Winfield$^{4}$, R.~Wood$^{8}$, P.~Woods$^{10}$ and A.~Yeremin$^{23}$}

\address{
$^1$ INTE-DFEN, Universitat Polit\`ecnica de Catalunya, E-08028 Barcelona, Spain\\
$^2$ IFIC, CSIC-Universitat de Val\`encia, E-46071 Valencia, Spain\\ 
$^3$ Institute of Nuclear Research of Hungarian Academy of Sciences, Debrecen, Hungary\\
$^4$ GSI Helmholtzzentrum f\"ur Schweionenforschung GmbH, D-64291 Darmstadt, Germany \\
$^5$ Universidade de Santiago de Compostela, E-15782 Santiago de Compostela, Spain\\
$^6$ National Research Centre ``Kurchatov Institute'', Moscow, Russia\\
$^7$ Joint Institute for Nuclear Research, 141980 Dubna, Russia\\
$^8$ Department of Physics, University of Surrey, Guildford GU2 7XH, United Kingdom\\
$^9$ CIEMAT, Madrid, Spain\\
$^{10}$ University of Edinburgh,  EH9 3JZ Edinburgh, United Kingdom\\
$^{11}$ Justus-Liebig Universit\"at Giessen, D-35392 Giessen, Germany\\
$^{12}$ St.~Mary's University, Halifax, B3H 3C3 Nova Scotia, Canada\\
$^{13}$ Physik Department E12, Technische Universit\"at M\"unchen, D-85748 Garching, Germany\\
$^{14}$ Centro de Fisica Nuclear da Universidade de Lisboa, 169-003 Lisboa, Portugal\\
$^{12}$ Universidad de Sevilla, Seville, Spain\\
$^{16}$ ExtreMe Matter Institute, Darmstadt, Germany\\
$^{17}$ Department of Physics, Faculty of Science, University of Zagreb, 10000 Zagreb, Croatia\\
$^{18}$ National\! Superconducting\! Cyclotron\! Laboratory,\! Michigan\! State\! University,\! East\!\! Lansing,\! USA\\
$^{19}$ Joint Institute for Nuclear Astrophysics, USA\\
$^{20}$ Instituto Nazionale di Fisica Nucleare, Laboratori Nazionale di Legnaro, Italy\\
$^{21}$ University of Notre Dame, 46556 South Bend - Indiana, USA\\
$^{22}$ University of Tennessee, Knoxville, 37996 Tennessee, USA\\
$^{23}$ Flerov Laboratory, Joint Institute for Nuclear Research, Dubna, Russia\\
$^{24}$ Institut de Physique Nucl\'eaire d'Orsay, France\\
}



\date{\today}

\begin{abstract}
The $\beta$-delayed neutron emission probabilities of neutron rich Hg and Tl nuclei have been measured together with $\beta$-decay half-lives for 20 isotopes of Au, Hg, Tl, Pb and Bi in the mass region N$\gtrsim$126. These are the heaviest species where neutron emission has been observed so far. These measurements provide key information to evaluate the performance of nuclear microscopic and phenomenological models in reproducing the high-energy part of the $\beta$-decay strength distribution. In doing so, it provides important constraints to global theoretical models currently used in $r$-process nucleosynthesis.
\begin{description}
\item[PACS numbers]\begin{verbatim}23.40.-s,25.70.Mn,26.30.Hj,27.80+w,29.38.Db+ \end{verbatim}
\end{description}
\end{abstract}

\pacs{23.40.-s,25.70.Mn,26.30.Hj,27.80+w,29.38.Db+}

\maketitle

The rapid neutron capture ($r$) process of nucleosynthesis represents a
challenge and a constant source of motivation for theory, experiments and
observations. Two paradigms are involved from the theory side, namely, the
modeling of suitable cataclysmic astrophysical environments and the
modeling of the atomic nucleus. Both of them are required if we are to understand
the origin of heavy elements as they are observed in the oldest stars, our
solar system and the Galaxy~\cite{Sneden08}. Two potential candidates for the $r$-process site are: core-collapse supernovae,
and mergers of either two neutron stars or a neutron star and a black hole~\cite{Freiburghaus99, Arnould07, Thielemann11, Goriely11, Arcones11, Rosswog13,Wallner15}. However, we face fundamental difficulties in ascribing either of them as the site of the stellar $r$-process. The latest hydrodynamical simulations of
supernovae and neutrino-driven winds fail to produce heavy nuclei and, in
particular, the prominent third $r$-process abundance peak at A$\sim$195~\cite{Arcones13}. However, explosions driven
by magnetic fields and rotation may contribute in the early galaxy to the
synthesis of heavy elements~\cite{Winteler12,Nishimura15,Takiwaki14}. On the other hand, neutron star mergers seem to provide
suitable physical conditions for an $r$-process environment but it is still
a matter of discussion whether their contribution came early enough to explain
the $r$-process abundances observed in the oldest stars (see e.g. Ref.~\cite{Argast04,Matteucci14,Shen15,Wehmeyer15,vdV15}).

In terms of the nuclear structure an important step forward came from the improvements in the theoretical description of nuclear masses~\cite{Moller95,Duflo95,Pearson96,Goriely09}, as well as microscopic approaches to $\beta$-decay strength function: QRPA model with no T=0 proton-neutron pairing employing the ground-state description from the FRDM~\cite{Moller03}, continuum proton-neutron QRPA based on the self-consistent density functional theory (DF3+cQRPA)~\cite{Borzov03}, the interacting shell model~\cite{Zhi13} and the fully self-consistent covariant density functional theory~\cite{Marketin15}. These studies enabled the realization of the role of first-forbidden (FF) transitions in shortening the $\beta$-decay half-lives of neutron-rich nuclei near neutron shell closures, mainly around $N=50$, $N=82$ and $N=126$. One of the remaining major difficulties for $r$-process calculations, aimed at reproducing the solar-system abundance distribution lies in the fact, that the performance of nuclear models far from stability and their robustness under extrapolation have not been tested~\cite{Arcones11,Surman14,Mumpower15}. Whereas nuclear masses (neutron-separation energies) essentially determine the path of the $r$-process along the nuclear chart, $\beta$-decay rates determine the speed at which the $r$-process operates. Finally, the competition between neutron capture, $\beta$-decay and $\beta$-delayed neutron emission during the $r$-process freeze-out shapes the final abundance distribution. The position and width of the third $r$-process peak at $A \sim 195$ has been found to be particularly sensitive to both half-lives and neutron branching ratios of the involved precursor nuclei around $N=126$~\cite{Arcones11,Domingo13,Eichler15,Mumpower15}.

In this respect, the $\beta$-decay strength distribution, which depends on the quantum structure of mother and daughter nuclei is particularly sensitive to the details of the nuclear model. In particular, the transitions to highly excited states in the $Q_{\beta}$ window provide the most stringent test for microscopic calculations and determine the $\beta$-delayed neutron emission probability ($P_n$), that can be also computed by phenomenological models~\cite{McCutchan12,Miernik13}. Recent efforts to improve phenomenological models~\cite{Miernik13,McCutchan12} are still affected by uncertainties of about a factor of two, reaching in some cases much larger discrepancies. Furthermore, the absence of nuclear microscopic properties in this kind of analytical prescription casts doubts on their extrapolation power, an aspect which can be investigated when new experimental data are obtained far from the region of reference. On the other hand, similar discrepancies are encountered in microscopic and/or macroscopic nuclear models~\cite{Borzov11,Moller95} when compared with the available data. 

Experimentally $\beta$-delayed neutron emission has been measured for only about 230 nuclei~\cite{Pritychenko11} out of the typically more than 5000 nuclei involved in $r$-process network calculations. With the sole exception of $^{210}$Tl, for which a neutron branching ratio of 7($^{+7}_{-4}$)$\times$10$^{-5}$ has been reported~\cite{Rusinov57,Stetter61}, there exists no other measurement in the heavy mass region around $N=126$. Indeed, most of the measured neutron-emitting nuclei correspond to fission products around A$\sim$95 and A$\sim$138. $^{150}$La, with a neutron branching ratio of 2.7(3)\% is the heaviest known nucleus with a substantial amount of neutron emission~\cite{Reeder86}. 

The present article reports on the first measurement of several $\beta$-delayed neutron emitters beyond N=126. The experiment was carried out at the heavy ion research centre GSI (Germany) using the fragment separator (FRS) in combination with the BEta-deLayEd Neutron (BELEN) detector~\cite{Agramunt15}. Indeed, $r$-process waiting point nuclei for $N=126$ and their direct neighbours are still inaccessible at present RIB facilities due to the very small production cross sections, the limited primary beam intensities and the challenging background conditions. The next generation of RIB facilities~\cite{Fulton11} and advances in production techniques~\cite{Watanabe15} are called to change this situation in the near future. In this work we used the high primary beam energy attainable at GSI with the SIS18 synchrotron, in order to produce and identify reliably Au-Hg-Tl-Pb-Bi nuclei with $N\gtrsim126$. The latter could be synthesized in sufficient yield by means of the fragmentation of a 1~GeV/u primary $^{238}$U beam impinging, with an intensity of 2$\times$10$^9$~ions/pulse, on a 1.6~g/cm$^2$ Be target. Fragments of interest were separated and selected with the FRS using the B$\rho$-$\Delta$E-B$\rho$ method~\cite{Geissel92}. 

\begin{figure}[!htbp]
\includegraphics[width=0.45\textwidth]{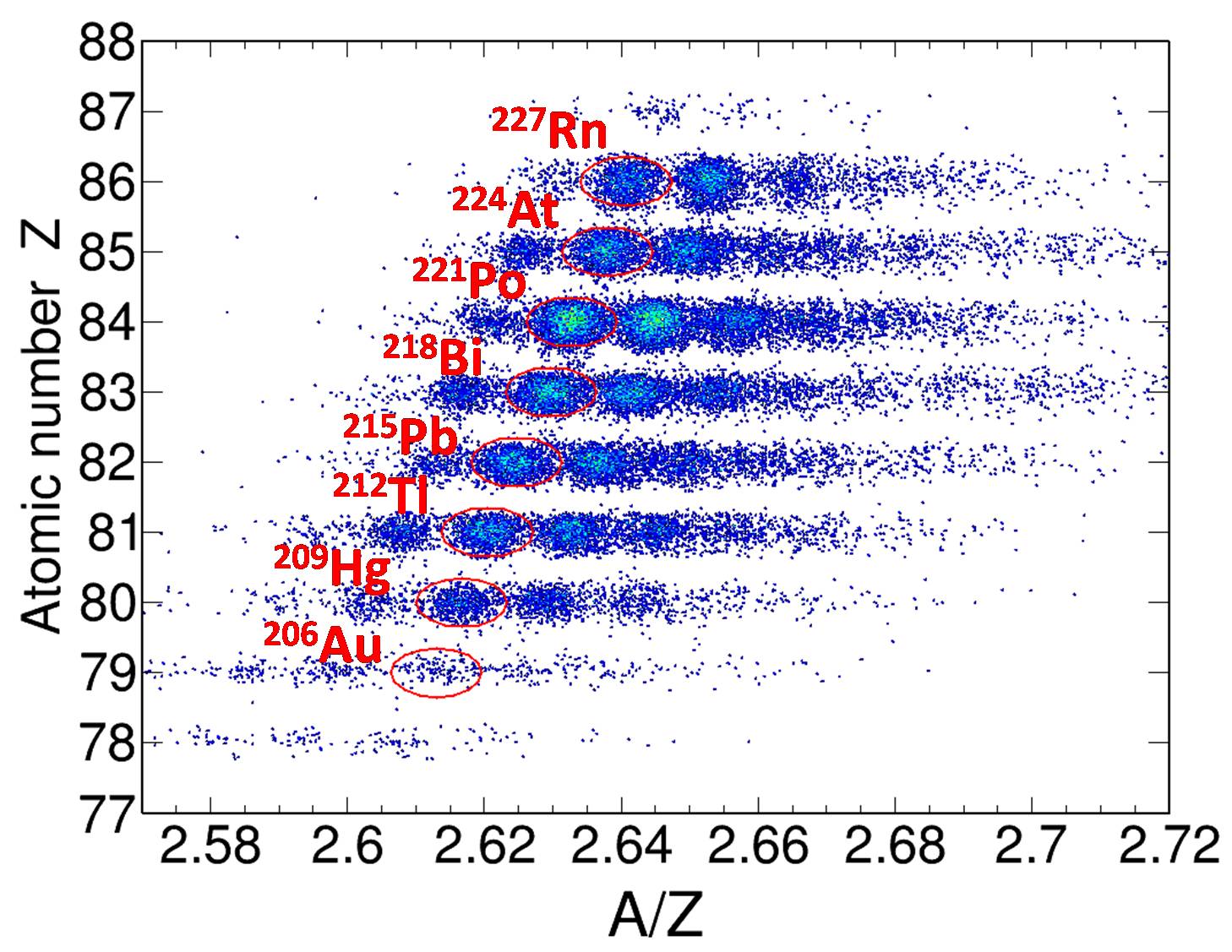}
\caption{\label{fig:pid} Ion identification diagram showing nuclear charge $Z$ as a function of the mass-over-charge ratio $A/Z$ as measured with the FRS tracking detectors (see text for details).}
\end{figure}

Ion tracking and identification was carried out by means of standard FRS tracking detectors, which comprised two time-projection chambers (TPC) at the intermediate focal plane (S2), another two at the final focal plane (S4), two ionization chambers at S4 and thin plastic scintillators at both S2 and S4 for measuring the time-of-flight (TOF). Ion identification was carried out on an event-by-event basis, by using the energy loss measured in the ionization chambers for charge determination and the measured TOF for determining the mass-over-charge (A/Z) ratio. Changes in the magnetic rigidity of the ions after the S2-degrader were used to identify and correct for charge states induced in the wedge~\cite{Caballero15}. A thin Nb foil was used between the two ionization chambers in order to induce electron stripping from possible H- or He-like charge states. By implementing ion-trajectory corrections enabled by the TPC-measurements, in combination with the TOF- and energy-loss information it was possible to determine the nuclear charge Z and the A/Z ratio. The final identification diagram is shown in Fig.\ref{fig:pid} for all the events accumulated in the experiment. An aluminium degrader with adjustable thickness was used at S4 in order to slow down ions and implant them into a stack of double-sided silicon strip detectors called SIMBA~\cite{Hinke12}. 

\begin{figure*}
\includegraphics[width=\textwidth]{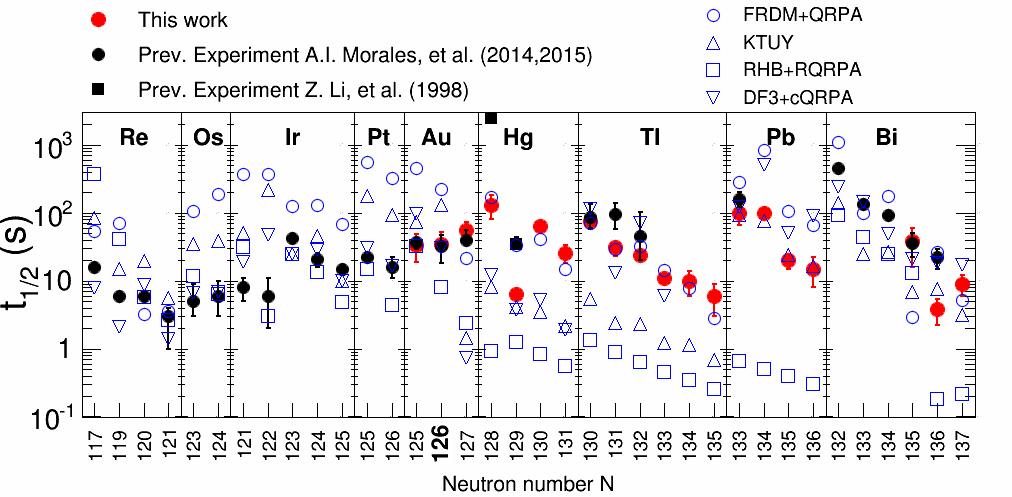}
\caption{\label{fig:thls} Experimental halflife values (solid symbols) on both sides of $N=126$ from previous experiments~\cite{Morales14a,Morales14b,Morales15} (bold circles) and \cite{Li98} (bold squares) and from the present work (red circles). Open circles show published theoretical half-lives~\cite{Moller03,Borzov03, Borzov11, Koura05, Marketin15} (see legend).}
\end{figure*}

Neutrons were detected using BELEN, which consisted of an array of 30 $^{3}$He-tubes embedded in a polyethylene matrix that served as moderator and shielding from surrounding background neutrons. \textsc{Geant4}~\cite{geant4} and MCNPX~\cite{mcnp} Monte Carlo simulations allowed us to determine the BELEN efficiency as a function of the neutron energy, which is nearly constant at a value of 40(2)\% from thermal neutrons up to 1 MeV, and decreases to 35\% at 5~MeV. The simulations were validated with dedicated measurements performed with a $^{252}$Cf-source. Data from BELEN were acquired using a digital data acquisition system~\cite{Agramunt13}, which was combined with the GSI local acquisition Multi-Branch System (MBS)~\cite{MBS}. The latter was used to acquire data from all tracking detectors and SIMBA. The MBS data acquisition was triggered by a scintillator detector at S4 upstream of the detector array and by high energy implant and low energy $\beta$-decay events in SIMBA. The digital data acquisition for BELEN was self-triggered. Always when there was an electrical signal from any of the $^{3}$He-tubes, its pulse-height and time-stamp were acquired. Both data acquisition systems generated time stamped data using a common clock. In this way, ion-implant versus $\beta$-decay time-correlations and ion-$\beta$-neutron time-correlations could be built with a sampling-resolution of 10~ns over an arbitrarily long time-window and in both forward (increasing) and backward (decreasing) time directions. By means of implant-$\beta$ and implant-$\beta$-neutron time-correlations in the forward (true and uncorrelated events) and backward (only uncorrelated events) directions, it was possible to carry out a binned maximum-likelihood analysis~\cite{Bernas90} and thus determine $\beta$-decay half-lives and $\beta$-delayed neutron branching ratios for the implanted nuclei. A correlation area of 3$\times$3~mm$^2$ centred around the implantation was used. 

\begin{figure}[!htbp]
\includegraphics[width=\columnwidth]{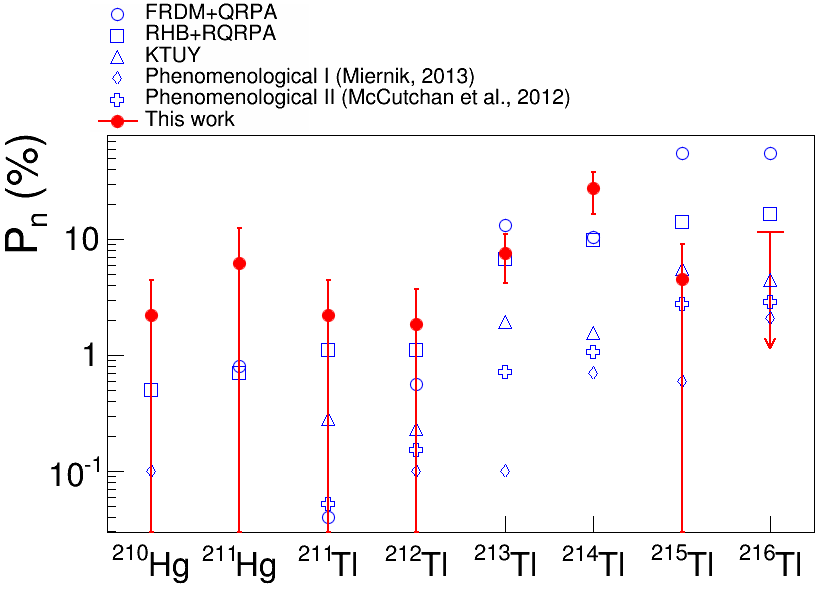}
\caption{\label{fig:pns} Measured neutron emission probability (\%) or upper limit ($^{216}$Tl) for the Hg- and Tl-isotopes (solid red circles and arrows). Theoretical predictions from Refs.~\cite{Moller03,Marketin15} and phenomenological models~\cite{Miernik13,McCutchan12} are shown with open symbols.}
\end{figure}

The results obtained for the half-lives and neutron branching ratios are reported in Table~\ref{table1} and displayed in Fig.~\ref{fig:thls} and Fig.~\ref{fig:pns}, respectively. 

\begin{table}[!htbp]
  \begin{tabular}{ccc|ccc}
    Isotope & t$_{1/2}$ (s) & P$_n$ (\%) & Isotope & t$_{1/2}$ (s) & P$_n$ (\%) \\
    \hline
    $^{204}$Au & 34(15) &  &$^{214}$Tl & 11(2) & 27(11)\\
    $^{205}$Au & 35(17) &  & $^{215}$Tl & 10(4) & 4.6(4.6)\\
    $^{206}$Au & 56(17) &  & $^{216}$Tl & 6(3) & $<11.5$\\
    $^{208}$Hg & 132(50) && $^{215}$Pb &  98(30)& \\
    $^{209}$Hg & 6(1) & &$^{216}$Pb &  99(12)& \\
    $^{210}$Hg & 64(12) & 2.2(2.2)&$^{217}$Pb &  20(5)& \\
    $^{211}$Hg & 26(8) & 6.3(6.3)&$^{218}$Pb &  15(7)& \\
    $^{211}$Tl & 76(18) & 2.2(2.2)&$^{218}$Bi &  38(22)& \\
    $^{212}$Tl & 31(8) & 1.8(1.8)&$^{219}$Bi &  3.8(1.6)& \\
    $^{213}$Tl & 24(4) & 7.6(3.4)&$^{220}$Bi &  9(3)& \\
  \end{tabular}
\caption{\label{table1}Implanted isotopes and measured half-lives and neutron branching ratios.}
\end{table}

In order to have a complete picture on both sides of the neutron shell closure in Fig.~\ref{fig:thls} we have included previously published results from another recent work~\cite{Benzoni12,Morales14a,Morales14b,Morales15}, as well as half-life values reported for $^{208,209}$Hg in Ref.~\cite{Li98}. The new results comprise twenty half-life measurements in the N$\gtrsim$126 region, nine of them reported for the first time. With the exception of $^{212}$Tl and $^{219}$Bi an excellent agreement is found with the recently published work of Morales \textit{et al.}~\cite{Benzoni12,Morales14a, Morales14b, Morales15}. Previously published half-lives for $^{208,209}$Hg~\cite{Li98} are much longer than the values reported here. A recent experiment at CERN-ISOLDE~\cite{Zsolt15} also concluded that the half-life of $^{208}$Hg is less than few minutes, and much shorter than the published value.
Thus, we use the recent half-life results to test the reliability of global theoretical models on both sides of $N=126$. The FRDM+QRPA~\cite{Moller03} model was until recently the only theory available over the full network of nuclei involved in $r$-process calculations. Presently, two additional complete $\beta$-decay calculations are available: The KTUY model, based on the gross theory with improved even-odd and shell-terms~\cite{Tachibana90,Koura05}, and the proton-neutron relativistic quasiparticle phase approximation based on the Hartree-Bogoliubov model (RHB+RQRPA)~\cite{Marketin15}. As already pointed out in Ref.~\cite{Morales14b}, in the $N \leq 126$ mass region the measured half-lives are on average a factor of 20 lower than FRDM+QRPA predictions. This effect has been explained on the basis of detailed calculations~\cite{Borzov06,Suzuki12,Zhi13}, which reveal an increasing contribution of FF transitions toward higher $Z$-values. This occurs because the Gamow-Teller (GT) transitions get progressively Pauli blocked by the filling of the h$_{11/2}$ proton orbital, which reduces contributions from allowed $\nu$h$_{9/2}$-$\pi$h$_{11/2}$ transitions. This behaviour in the $N \leq 126$ mass region seems to be properly reproduced by the latest global calculation RHB+RQRPA, where the average discrepancy is only a factor of three with respect to the experimental results. The performance of the highly parametrized KTUY scheme is worse, with half-lives which are, on average, a factor of six higher.

The situation, however, is reversed beyond $N=126$. In this region FRDM+QRPA predicts half-lives which are, on average, in good agreement with the experimental values. This is particularly true for the chain of Tl isotopes, where only minor discrepancies between FRDM+QRPA and experiment can be observed over the mass range from 211 up to 216. The decay systematics of the Tl isotopic chain, from $^{211}$Tl up to $^{213}$Tl, have been thoroughly investigated on the basis of half-life and spectroscopy measurements~\cite{Benzoni12,Morales14a,Morales14b}. With the results of the present work, this can be extended three mass units further, i.e. to $^{214,215,216}$Tl. Our new results confirm the systematic behaviour described in \cite{Morales14b} for the Tl isotopic chain, which is consistent with DF3+cQRPA calculations published for $^{211-214}$Tl\cite{Borzov03} (Fig.~\ref{fig:thls}). This was also interpreted~\cite{AlDahan09,Benzoni12,Morales14a,Morales14b} as an indication of the predominance of allowed $\nu i_{11/2} \rightarrow \pi i_{13/2}$ GT-transitions in the $N > 126$, $Z \leq 82$ mass-region ``south-east'' from $^{208}$Pb. This explanation also holds for the $\beta$-decay of $^{206}$Au, where our result is in agreement with the value reported recently~\cite{Morales15}. 
Whether the anomalously high occupation of the $\nu i_{11/2}$ orbital is due to a weakening of the spin-orbit field caused by the tensor force~\cite{Goddard13} or to the recently proposed three-body force mechanism~\cite{Nakada15a,Nakada15b,Gottardo12}, remains an open question which calls for further specific experiments.

KTUY reproduces excellently the half-lives measured for the lead isotopes (Fig.~\ref{fig:thls}), but the remainder of the half-lives in the $N>126$ region are underestimated, on average, by almost one order-of-magnitude.

The good performance of the RHB+RQRPA framework in the $N \leq 126$ region seems to vanish beyond the neutron shell closure for $N>126$. However, this seems to be a problem inherent to all global calculations presently available, where a fair performance in the $N<126$ region (KTUY and RHB+RQRPA) implies a poor behabiour beyond the shell closure ($N>126$) and vice-versa (FRDM+QRPA). In particular, the RHB+RQRPA predicted half-lives for $N>126$ nuclei are, on average, almost a factor of sixty lower than the experimental results. This may be ascribed to the fact, that state-of-the-art self-consistent nuclear models are not expected to provide accurate single-particle energies, particularly near closed-shell nuclei, where the $Q_{\beta}$ is not large and the strength may be dominated by only one or few transitions. In addition, the strength of the T=0 proton-neutron pairing is included in Ref.~\cite{Marketin15} using a single and rather simple parameterization~\cite{Niu13}.
In the DF3+cQRPA~\cite{Borzov03}, the effective mass m$^*$=1 and a moderate mass-independent strength of the T=0 proton-neutron pairing is used, resulting in a more satisfactory description of the Tl to Bi isotopes. Interestingly, in the FRDM+QRPA where the T=0 pairing is absent, the description of the semi-magic Z=82$\pm$1 isotopes, for which it is not that important, is better compared to other isotopic chains.

Additional constraints when comparing microscopic models with experiment can be found in the neutron-branching ratios of the Tl-isotopes, which are shown in Fig.~\ref{fig:pns}. The discrepancies found between the RHB+RQRPA predictions and the measured Tl-decay half-lives (Fig.~\ref{fig:thls}) are at variance with the quite good agreement that is found with the measured neutron branching ratios. As the GT and FF decays are treated in a single RQRPA framework, this may reflect that the integral of GT+FF strength is overestimated both in $Q_{\beta n}$ and $Q_{\beta}$ windows.
The good agreement between the experimental half-lives and the FRDM+QRPA predictions for the entire Tl-isotopic chain (Fig.~\ref{fig:thls}) is in contrast with the factor of five discrepancy found for the FRDM+QRPA predicted branching ratios of $^{215,216}$Tl. Such inconsistency might be ascribed to the QRPA description of the GT and statitical treatment of the FF transitions.
The KTUY predicted neutron branching ratios seem reasonable, but with a clear tendency to underestimate them.

Finally, Fig.~\ref{fig:pns} also shows a comparison between measured neutron branching ratios and phenomenological models~\cite{McCutchan12,Miernik13}. Agreement is found only with the neutron branching ratios measured for $^{210,211}$Hg and $^{211,212,215,216}$Tl, which have a rather large uncertainty. This feature points to an accidental coincidence rather than a predictive power. Again, more measurements will be important in order to extend the parameterized models to the heavy mass region.

In summary, the present work reports experimental results for nine new $\beta$-decay half-lives and eight new $\beta$-delayed neutron emission probabilities (or upper limits) in the heavy mass region beyond $N=126$. These new data have allowed us to get an insight into the $\beta$-decay performance of the only three nuclear models that are applicable over the full nuclear landscape, FRDM+QRPA~\cite{Moller03}, KTUY~\cite{Koura05} and RHB+RQRPA~\cite{Marketin15}. 
One can conclude that, presently, there exist no global model which provides satisfactory $\beta$-decay half-lives and neutron branchings on both sides of the $N=126$ shell closure. The RHB+RQRPA framework predicts fairly well the $\beta$-decay half-lives in the $N\leq 126$ region (although not beyond the shell-closure) as well as the neutron emission probability, at least in the region where new data are reported. It is worth noting that in the $N\leq 126$ region, where the new RHB+RQRPA model and the previous self-consistent study~\cite{Borzov03,Borzov06,Borzov11,Borzov14} perform better, is more relevant than the nuclei beyond the neutron shell closure because it involves all the species produced during the $r$-process freeze-out. Thus, the new experimental data support the rather good agreement predicted in Ref.~\cite{Marketin15} for the $r$-process abundances around the third $r$-process peak at $A=195$. 

The next important step toward a fully self-consistent description of the $\beta$-decay rates including nuclear deformation has been done within the Finite Amplitude Method (FAM)~\cite{Mustonen14,Mustonen15}. In this respect, a remaining challenge of prime importance for the QRPA is the impact of the 2p-2h configurations on the $\beta$-decay rates and $P_{n}$ values~\cite{Marketin12,Severyukhin13,Gambacurta15}. 
Further measurements of these two gross properties of $\beta$-decay and possibly spectroscopic measurements in the heavy mass region around A$\sim$200 are highly desirable in order to further improve theoretical models and $r$-process calculations. 

\begin{acknowledgments}
C.D.P. acknowledges helpful discussions with A.I.~Morales. This work was supported by the Spanish \textit{Ministerio de Economia y Competitividad} under Grants No. SEV-2014-0398, FPA2011-24553, FPA2011-28770-C03-03, FPA2014-52823-C2-1-P, AIC-D-2011-0705, the Spanish Nuclear Security Council (CSN) under a grant of Catedra Argos, by the German Helmholtz Association via the Young Investigators Grant VH-NG 627 and the Nuclear Astrophysics Virtual Institute (VH-VI-417), and by the German \textit{Bundesministerium f\"ur Bildung und Forschung} under No. 06MT7178 / 05P12WOFNF. UK authors acknowledge the support of the UK Science \& Technology Facilities Council. 

\end{acknowledgments}

\appendix

\bibliography{bibliography_s410}

\end{document}